# Giant Spin-valley Polarization and Multiple Hall Effect in Functionalized Bi Monolayers

*Tong Zhou[1,2], Jiayong Zhang[1,3], Hua Jiang[4], Igor Zutic[2], and Zhongqin Yang[1,5,*]*

[1]State Key Laboratory of Surface Physics and Key Laboratory for Computational Physical Sciences (MOE) & Department of Physics, Fudan University, Shanghai 200433, China

[2]Department of Physics, University at Buffalo, State University of New York, Buffalo, New York 14260, USA

[3]Jiangsu Key Laboratory of Micro and Nano Heat Fluid Flow Technology and Energy Application, School of Mathematics and Physics, Suzhou University of Science and Technology, Suzhou 215009, Jiangsu, China

[4]College of Physics, Optoelectronics and Energy, Soochow University, Suzhou 215006, China

[5]Collaborative Innovation Center of Advanced Microstructures, Fudan University, Shanghai 200433, China

*Address correspondence to: zyang@fudan.edu.cn

## ABSTRACT:

Valleytronic materials, characterized by local extrema (valley) in their bands, and topological insulators have separately attracted great interest recently. However, the interplay between valleytronic and topological properties in one single system, likely to enable important unexplored phenomena and applications, has been largely overlooked so far. Here, by combining a tight-binding model with first-principles calculations, we find the large-band-gap quantum spin Hall effects (QSHEs) and valley Hall effects (VHEs) appear simultaneously in the Bi monolayers decorated with halogen elements, denoted as $Bi_2XY$ (X, Y = H, F, Cl, Br, or I). A staggered exchange field is introduced into the $Bi_2XY$ monolayers by transition metal atom (Cr, Mo, or W) doping or $LaFeO_3$ magnetic substrates, which together with the strong SOC of Bi atoms generates a time-reversal-symmetry-broken QSHE and a huge valley splitting (up to 513 meV) in the system. With gate control, QSHE and anomalous charge, spin, valley Hall effects can be observed in the single system. These predicted multiple and exotic Hall effects, associated with various degrees of freedom of electrons, could enable applications of the functionalized Bi monolayers in electronics, spintronics, and valleytronics.

**INTRODUCTION**

Tailoring valley degrees of freedom offers fascinating opportunities to realize novel phenomena and emerging applications, often referred to as valleytronics[1-4]. While valley effects have been studied for decades in materials such as silicon[5], diamond[6], AlAs[7], and graphene[8-11], despite the effort to emulate the better known manipulation of spin and spintronic applications[12], the related success has been modest[2]. Often the progress in harnessing the valley degrees of freedom was limited not by the lack of ideas[13], but by the material properties, including the small valley polarization and a weak spin-orbit coupling (SOC) inherent to graphene[1,8,11]. The resurgence of interest in valley effects was recently spurred by the discovery of monolayer (ML) transition metal dichalcogenides (TMDs) with broken inversion symmetry and strong SOC[14-22]. A hallmark of ML TMDs is their valley-spin coupling which leads to a valley-dependent helicity of optical transitions[23-25] as well as important implications for transport, such as the discovery of the valley Hall effect (VHE)[4]. Lifting the degeneracy between the valleys K and K' to generate the valley polarization was identified as the key step in manipulating valley pseudospin degrees of freedom[15-22]. Common approaches were focused either on optical pumping by circularly polarized light[23-25] or very large magnetic fields required by a small Zeeman splitting of ~0.1 meV/T[26,27]. Instead of these external methods to realize valley polarization that could be impractical or limited by the carrier lifetime[3], transition metal (TM) doping[16-18] or magnetic proximity effects[19-22] have been found effective ways to get permanent valley polarizations in TMDs. Since topological properties are very important for materials[1], the coupling of the topological behaviors and valley polarizations in the valleytronic materials may give rise to new physics and applications. However, the widely studied valley-polarized material TMDs have topologically trivial band gaps[16-22]. The combination of topological and valleytronic has been rarely explored up to the present. It would be very interesting to search for such topologically nontrivial valleytronic materials and explore their valley-dependent transport properties.

Since the topological properties and transverse velocities of the electrons in the VHEs are closely related to the SOC strength of the system, it is desirable to search for and to explore the valley related phenomenon in large SOC material systems. Very strong SOC can be generally induced in materials containing heavy elements, such as bismuth, and drives the appearance of the topologically nontrivial states in the Bi-related systems[28,29]. In experiments, the buckled ML structure of Bi (111) with a hexagonal lattice has been successfully synthesized and characterized[28]. A nontrivial quantum spin Hall effect (QSHE) band gap of 0.8 eV was observed experimentally in ML Bi on the top of a SiC substrate[28]. With the hydrogen (H)[30,31], halogens (F, Cl, Br, I)[31] or methyl ($CH_3$)[32] saturating $p_z$ orbitals of Bi atoms, the buckled Bi (111) ML forms a hexagonal flat geometry $Bi_2X_2$ (X= H, F, Cl, Br, I, or $CH_3$), of which atomic structures are predicted to be stable at up to 600 K[31]. A giant SOC strength, producing global nontrivial band gaps up to 1.03 eV, was reported in the $Bi_2X_2$ system[31,32]. These $p_x$ and $p_y$ orbital-induced large SOC should be beneficial for coupled spin and valley physics. Due to the presence of inversion symmetry in the crystal structure of $Bi_2X_2$, no valley-related phenomenon emerges in the previous work[28,30-32].

In this work, by focusing on strong SOC in ML Bi (111)-based systems, we use tight binding (TB) model and *ab-initio* studies to demonstrate materials design of topological effects which rely on valley-dependent Berry curvature. Large nontrivial band gaps (from 0.891 eV to 1.256 eV) are obtained at the two valleys in ML Bi (111) with different chemical decorations on the two surfaces of the MLs, forming $Bi_2XY$ (X, Y = H, F, Cl, Br, or I and X≠Y) structures, which provide a platform for fabricating the wide-frequency valley-light devices. To break the degeneracy of the K and K' valleys, we induce a staggered exchange field $\Delta M = M_A - M_B$ into the ML $Bi_2XY$ with the assumption of $M_A > M_B$, where $M_A$ is the magnetic exchange field in the A sublattice of the ML Bi and $M_B$ is the magnetic exchange field in the B sublattice. The TB model calculations show that when SOC is larger than both $M_A$ and $M_B$, the time reversal symmetry broken (TRSB) QSH states and obvious spin-valley polarizations emerge in the system. With electron (hole) doping, the spin-down electrons (spin-up holes) produce a spin-valley polarized net transverse current, giving rise to spin-valley polarized anomalous valley Hall effects (AVHEs), also called anomalous charge/spin/valley Hall effects. Thus, with gate control, multiple Hall effects including QSHEs and anomalous charge/spin/valley Hall effects can be manipulated in the single system. To carry out these effects in experiments, we propose two possible schemes based on *ab-initio* calculations. 1) A functionalized ML $Bi_2XY$ with transition metal (TM = Cr, Mo, or W) atoms doped, in which the valley splitting can be giant, with a maximum value of 513 meV. 2) A heterostructure of ML $Bi_2H$ deposited on a $LaFeO_3$ (111) surface, where a valley splitting is about 78 meV. The predicted multiple Hall effects associated with multiple degrees of freedom of electrons in functionalized ML Bi pave a brand new way to electronics, spintronics, and valleytronics of two-dimensional materials.

## RESULTS AND DISCUSSION

The structure of ML $Bi_2XY$ (X, Y = H, F, Cl, Br, or I) is shown in Figs. 1a and b, where the Bi atoms construct a quasi-planar honeycomb lattice, the X atoms bond the upper surface Bi atoms, and the Y atoms bond the lower surface Bi atoms. When X = Y, the $Bi_2XY$ structures form the ML $Bi_2X_2$ (X = H, F, Cl, Br, or I), which has been predicted to be stable up to 600 K and show QSH effects[31]. The ML $Bi_2X_2$ has inversion symmetry with point group of $D_{3d}$[31]. The Bi $p_z$ orbitals are saturated by the later added X atoms. The strong SOC from Bi $p_x$ and $p_y$ orbitals near the Fermi level ($E_F$), thus, opens large nontrivial band gaps at the K and K' valleys[30-32]. The calculated local band gaps of $Bi_2X_2$ (X = H, F, Cl, Br, or I) are found in the range from 1.160 eV to 1.306 eV as shown in Table S1 and Fig. S1 in Supplementary Information. A large SOC in the ML $Bi_2X_2$ could be employed for the potential valleytronic applications after the inversion symmetry of the system is broken.

To explore the valley contrasting physics in the functionalized Bi systems, a TB model is built for the ML $Bi_2XY$. We adopt the spherical harmonic functions $|\phi_+\rangle = -\frac{1}{\sqrt{2}}(p_x+ip_y)$ and $|\phi_-\rangle = \frac{1}{\sqrt{2}}(p_x-ip_y)$ together with the spin $\{\uparrow,\downarrow\}$ as the basis. The TB Hamiltonian of the ML $Bi_2X_2$ is first discussed. Under the basis of $\Phi_i = \{|\phi_+\rangle, |\phi_-\rangle\} \otimes \{\uparrow,\downarrow\}$, the Hamiltonian ($H_1$) can be written in a sum of the nearest neighbor hopping ($H_0$) and on-site intrinsic SOC terms ($H_{SO}$)[33]:

$$H_1 = H_0 + H_{SO}, \qquad (1)$$

$$H_0 = \sum_{i \in A} \sum_{\delta=1,2,3} \Phi_i^+ T_\delta \Phi_{i+\delta} + H.C., \quad (2)$$

$$H_{SO} = \lambda_{so} \sum_i \Phi_i^+ \tau_z \otimes \sigma_z \Phi_i. \quad (3)$$

In Eqs. (2) and (3), $\Phi_i$ represents the annihilation operator on site i. $\tau_z$ and $\sigma_z$ in the equations indicate the Pauli matrices acting on orbital $\{|\phi_+\rangle, |\phi_-\rangle\}$ and spin $\{\uparrow, \downarrow\}$ spaces, respectively. $\delta_1 = (1,0)$, $\delta_2 = (-\frac{1}{2}, \frac{\sqrt{3}}{2})$, and $\delta_3 = (-\frac{1}{2}, -\frac{\sqrt{3}}{2})$ are the three vectors to the nearest neighbor sites. The $T_\delta$ in Eq. (2) takes the forms of

$$T_{\delta_1} = \begin{bmatrix} t_1 & t_2 \\ t_2 & t_1 \end{bmatrix} \otimes \sigma_0, \quad T_{\delta_2} = \begin{bmatrix} t_1 & z^2 t_2 \\ z t_2 & t_1 \end{bmatrix} \otimes \sigma_0, \quad T_{\delta_3} = \begin{bmatrix} t_1 & z t_2 \\ z^2 t_2 & t_1 \end{bmatrix} \otimes \sigma_0, \quad (4)$$

with $z = \exp(\frac{2}{3}i\pi)$. $t_1$ and $t_2$ in Eq. (4) represent the hopping amplitudes. For the ML Bi$_2$X$_2$ there are Dirac points existing at the K and K' points without the SOC, as shown in Fig. 1c. When SOC is considered, nontrivial band gaps of 2λso are opened at the K and K' points shown in Fig. 1d, realizing a QSH insulator, in agreement with the *ab-initio* calculations (see Fig. S1 in Supplementary Information) and previous studies[30-32]. The local band gap opened at the Dirac point K (K') is a result of the first-order relativistic effect related to $p_x$ and $p_y$ orbitals of Bi elements[31]. Thus, these band gaps are giant and robust (1.160 ~ 1.306 eV, see Table S1 in Supplementary Information).

To break the inversion symmetry of the Bi$_2$X$_2$ system, we induce a staggered potential ($H_U$) between the sublattices A and B, equivalently forming the Bi$_2$XY system. Simultaneously, the Rashba SOC[12] exists in the system due to the different decorations of X and Y. The $H_U$ can be written as $H_U = U \sum_i \mu_i \Phi_i^+ \tau_0 \otimes \sigma_0 \Phi_i$, where $\mu_i = 1$ (−1) for the A (B) sublattice and both $\sigma_0$ and $\tau_0$ are 2×2 unitary matrices. The Rashba SOC $H_R$ can be written as $H_R = \sum_{i \in A} \sum_{\delta=1,2,3} \Phi_i^+ T_{R\delta} \Phi_{i+\delta}$[33]. The resulting TB Hamiltonian for the ML Bi$_2$XY is thus:

$$H_2 = H_0 + H_{SO} + H_U + H_R, \quad (5)$$

where

$$T_{R\delta_1} = -i\begin{bmatrix} \lambda_R & \lambda_R' \\ \lambda_R' & \lambda_R \end{bmatrix} \otimes \sigma_y, \quad T_{R\delta_2} = i\begin{bmatrix} \lambda_R & z^2\lambda_R' \\ z\lambda_R' & \lambda_R \end{bmatrix} \otimes (\frac{\sqrt{3}}{2}\sigma_x + \frac{1}{2}\sigma_y), \quad T_{R\delta_3} = i\begin{bmatrix} \lambda_R & z\lambda_R' \\ z^2\lambda_R' & \lambda_R \end{bmatrix} \otimes (-\frac{\sqrt{3}}{2}\sigma_x + \frac{1}{2}\sigma_y), \quad (6)$$

with $z = \exp(2/3i\pi)$. $\lambda_R$, $\lambda_R'$ reflect the Rashba SOC between different orbitals of nearest-neighbor sites. In general, the relation of $\lambda_R$ and $\lambda_R'$ follows tendency of the $t_1$ and $t_2$. We assume $\lambda_R' = \lambda_R t_2/t_1$ for simplicity and to reduce the number of the independent parameters. The Rashba SOC is important in some systems, which can induce very interesting effects. For example, it opens the topologically nontrivial gap of quantum anomalous Hall effects in graphene system[34,35]. However, in the Bi$_2$XY system, the Rashba SOC is small compared to intrinsic SOC and staggered potential, which will be shown in the following band fitting. Without the SOC, the bands of the Hamiltonian $H_0 + H_U$ are plotted in Fig. 1e. The spin-up and spin-down bands are degenerate at the K and K' points. Because of the staggered potential, the Dirac points disappear and two trivial band gaps with the value of *2U* emerge at the K and K' points. With the consideration of the SOC and $\lambda_{SO} > U$, the topologically nontrivial band gaps are opened at the K and K' points (Fig. 1f). The SOC together with staggered potential lifts the spin degeneracy of energy bands and makes the system have a peculiar spin-valley coupling, as shown in Fig. 1f. An obvious spin splitting appears at both the valence and conduction bands with opposite spin moments in the two valleys. An obvious spin splitting appears at both the valence and conduction bands with opposite spin moments in the two valleys. The calculated Berry curvatures are opposite at different valleys (Fig. 1f), providing an effective magnetic field. Such a magnetic field not only defines the optical selection rules, but also generates an anomalous velocity for the charge carriers. Thus, the valley Hall effect exists in the system. However, in the presence of time-reversal symmetry, the energy bands are still valley degenerate when SOC is considered. The opposite Berry curvatures and spin moments at the two valleys give rise to both the valley and spin Hall effects, without net transverse charge Hall current.

In the real ML Bi (111), the staggered potential can be acquired with different chemical decorations on the upper and lower ML surfaces, namely forming the $Bi_2XY$ (X, Y = H, F, Cl, Br, or I) structure. The calculated staggered potentials are found in the range from 6 meV to 300 meV, as shown in Table S2 and Fig. S2 in Supplementary Information. With the SOC, large nontrivial band gaps from 0.891 eV to 1.256 eV (see Table S1 and Fig. S3 in Supplementary Information) are opened in the two valleys, giving rise to quantum spin Hall effect and valley Hall effect. These band gaps of the ML $Bi_2XY$ not only give us an opportunity to generate valley polarizations with different wavelength of light[23-25], but also provide a platform for fabricating the wide frequency valley-light emitting diodes[36]. As an example, we now analyze the case of the ML $Bi_2HF$, where the H atoms bond with the lower Bi atoms and the F atoms bond with the upper Bi atoms. We find that the lattice constant of this ML $Bi_2HF$ is 5.49 Å after a full relaxation. The calculated bond lengths of Bi-H and Bi-F are 1.84 Å and 2.08 Å, respectively. The energy bands for the ML $Bi_2HF$ without SOC are displayed in Fig. 2a. The band gaps at the K and K' points are found to be 141 meV. Comparing with Fig. 1c, it can be inferred that the staggered potential *2U* in the ML $Bi_2HF$ is about 141 meV, large enough to break the inversion symmetry in the system. When the SOC is turned on, very large nontrivial band gaps of 1.16 eV are opened around the $E_F$ at the K and K' points. The orbital-resolved band structures for $Bi_2HF$ are shown in Fig. S4. It is clear that the bands around the $E_F$ at the two valleys are dominated by the $p_x$ and $p_y$ orbitals of Bi atoms, indicating the constructed TB model can describe the $Bi_2HF$ well. By fitting TB bands to the *ab-initio* results (shown in Fig. S4), the obtained TB parameters for the ML $Bi_2HF$ are $t_1$ = 0.73 eV, $t_2$ = 1.06 eV, 2U = 0.14 eV, $\lambda_{SO}$ = 0.61 eV, and $\lambda_R$ = 0.022 eV. Very obvious spin polarization is also observed in the highest valence bands (HVBs) and the lowest conduction bands (LCBs).

To identify the QSH effect and VHE emerging in the ML $Bi_2HF$, the spin Berry curvature $\Omega^s(\mathbf{k})$[14] and Berry curvature $\Omega(\mathbf{k})$[37] are calculated as follows:

$$\Omega^s(\mathbf{k}) = \sum_n f_n \Omega^s_n(\mathbf{k}), \tag{7}$$

$$\Omega_n^s(\mathbf{k}) = -2\,\text{Im}\sum_{m\neq n} \frac{\langle\psi_{n\mathbf{k}}|j_x^s|\psi_{m\mathbf{k}}\rangle\langle\psi_{m\mathbf{k}}|v_y|\psi_{n\mathbf{k}}\rangle\hbar}{(E_m - E_n)^2}, \quad (8)$$

$$\Omega(\mathbf{k}) = \sum_n f_n \Omega_n(\mathbf{k}), \quad (9)$$

$$\Omega_n(\mathbf{k}) = -2\,\text{Im}\sum_{m\neq n} \frac{\langle\psi_{n\mathbf{k}}|v_x|\psi_{m\mathbf{k}}\rangle\langle\psi_{m\mathbf{k}}|v_y|\psi_{n\mathbf{k}}\rangle\hbar^2}{(E_m - E_n)^2}. \quad (10)$$

In Eqs. (7) - (9), $E_n$ is the eigenvalue of the Bloch functions $|\psi_{n\mathbf{k}}\rangle$, $f_n$ is the Fermi-Dirac distribution function at zero temperature, and $j_x^s$ is the spin current operator defined as $(s_z v_x + v_x s_z)/2$, where $v_x$ and $v_y$ are the velocity operators and $s_z$ is the spin operator. The obtained spin Berry curvature $\Omega^s(\mathbf{k})$ along the high-symmetry lines is plotted in Fig. 2c. By integrating $\Omega^s(\mathbf{k})$ over the first Brillouin zone (BZ), we obtain the spin Chern number Cs = 1, proving that the ML Bi$_2$HF is a QSH insulator. As shown in Fig. 2d, the calculated $\Omega(\mathbf{k})$ is sharply peaked in the valley region, with opposite signs for K and K', in agreement with Fig. 1f. The distribution of $\Omega(\mathbf{k})$ the indicates the valley Hall effect occurring in the system. Namely, when the ML Bi$_2$XY channel is biased, electrons from different valleys experience opposite Lorentz-like forces and so move in opposite directions perpendicular to the drift current.

To employ the valley degree of freedom, valley splitting ($\Delta_{KK'}^V$) needs to be introduced, which can be quantified by the energy difference between the topmost valence bands at K ($E_K^V$) and K' ($E_{K'}^V$) valleys, expressed as $\Delta_{KK'}^V = E_K^V - E_{K'}^V$ [16-21]. In this regard, the principle challenge of using ML Bi$_2$XY as a valleytronic material is to break the degeneracy between the two prominent K and K' valleys, protected by time reversal symmetry. As in TMDs, some strategies of using external fields such as optical pumping[23-25], electric field[9], and magnetic field[26,27] are supposed to induce the valley polarizations in the ML Bi$_2$XY. Importantly, for nonvolatile devices, magnetic doping[16-18], proximity coupling with magnetic substrates[19-22,38] or spontaneous valley polarization[15,39] could be desirable to provide intrinsic and robust lifting of the valley degeneracy[17,21,22]. Since the spin and valley degrees of freedom are strongly coupled in the ML Bi$_2$XY, spin polarization could be exploited to induce valley polarization. Here, we introduce a staggered exchange field $H_M = M_{A(B)}\sum_{i\in A(B)} \Phi_i^+ \tau_0 \otimes \sigma_z \Phi_i$ into the system, where $M_{A(B)}$ is the local magnetic exchange field in the A (B) sublattice. Such a staggered exchange field has been proven to transform the bands at K and K' points in the silicence[40] and functionalized ML Sb[33,41], leading to so-called quantum spin quantum anomalous Hall (QSQAH) effects[33,40,41]. The staggered exchange field is thus suitable to produce the desired spin-valley polarizations in the ML Bi$_2$XY. The TB Hamiltonian of the ML Bi$_2$XY with staggered exchange field can be written as:

$$H_3 = H_0 + H_{SO} + H_U + H_M + H_R. \tag{11}$$

The schematic bands for $H_3$ without and with SOC are plotted as Figs. 3a and b, respectively. With the staggered exchange field but without SOC, the bands are spin polarized, giving spin splittings of $M_A + M_B - 2U$ and $M_A + M_B + 2U$ for LCBs and HVBs, respectively. When $M_A + M_B$ is larger than $2U$, the bands are inverted near the $E_F$ as shown in Fig. 3a. However, the two valleys are still degenerate. Since the SOC of Bi atoms, primarily coming from the $p_x$ and $p_y$ orbitals, is very large, $\lambda_{SO} > M_A > M_B$ can be assumed in the TB model calculations. Under this condition, two local band gaps with different values are opened at the K and K' points, indicating a valley polarization in the system (Fig. 3b). To quantitatively analyze the sequence of the energy levels around the $E_F$ at the K and K' points, the Rashba SOC is neglected because it is small compared to intrinsic SOC and magnetic field. Analyzing the sequence of the energy levels around the $E_F$ at the two valleys from Fig. 3b (also see Fig. S5 and more details in Supplementary Information), we can obtain the local band gaps at the K and K' points with the values of $2\lambda_{SO} - 2M_A$ and $2\lambda_{SO} - 2M_B$, respectively. These band gaps give rise to an interesting time-reversal-symmetry-broken (TRSB) QSH state[42]. The calculated spin Berry curvatures (Fig. 3c) and edge states (Fig. 3d) are displayed to identify the interesting TRSB-QSH states. Thus, when the $E_F$ is in the nontrivial band gap ( $E_K^C > E_F > E_K^V$ ), TRSB-QSH edge states could be observed as shown in Fig. 3e.

For applications, static and large valley polarization is desirable. In the ML Bi$_2$XY, the charge carriers in the two valleys have opposite transverse velocities due to the valley degeneracy and opposite signs of the Berry curvatures and thus the total Hall conductivity vanishes because of time-reversal symmetry. With the staggered exchange field induced, the valley splitting is obtained. Considering the energy levels at two valleys (see more details in Supplementary Information) we can get the valley splitting for the valance bands ($\Delta^V_{KK'}$) is $\Delta M + 2U$ ($\Delta M = M_A - M_B$). Similarly, we can also obtain the valley splitting for the conduction bands ($\Delta^C_{KK'}$) of $\Delta M - 2U$. Clearly, the valley splittings in the functionalized ML Bi are determined by the staggered potential ($2U$) and exchange fields ($\Delta M$). Based on the valley splitting, one can manipulate the charge/spin/valley Hall effect with gate control. For example, when $E^V_K > E_F > E^V_{K'}$, the up-spin holes at the K valley produce a transverse current under an in-plane electric field, as shown in Fig. 3f. Remarkably, the transverse current is 100% spin polarized. Thus, the flux of the spin holes carries three observable quantities: charge, spin, and valley-dependent orbital magnetic moments corresponding, respectively, to anomalous charge, spin, and valley Hall effects. Similarly, when $E^C_{K'} > E_F > E^C_K$, spin-down electrons will produce a net transverse charge/spin/valley current, as shown in Fig. 3f. Therefore, we create intrinsic and robust valley polarization, instead of the using dynamic methods[23-25]. Considering QSHE also exists in the system when $E^C_K > E_F > E^V_K$, multiple Hall effects including TRSB-QSHE, anomalous charge/spin/valley Hall effects can be manipulated by gate voltage control in one single system as shown in Figs. 3e and 3f. Based on this multiple Hall effects control, we can flexibly manipulate the charge, spin, and valley degrees for transportation, which is crucial for spintronics and valleytronics.

In experiments, the staggered exchange fields may be induced by replacing the Y atoms with TM (Cr, Mo, W) atoms, forming Bi$_2$XTM (X = H, F, Cl, Br, or I; TM= Cr, Mo, or W) structures. To realize this proposal, we calculate the electronic structures of Bi$_2$HTM (TM= Cr, Mo, or W) monolayers by using *ab-initio* methods. The geometry and calculation details of Bi$_2$HTM (TM = Cr, Mo, or W) monolayers are given in Fig. S6 in Supplementary Information. For convenience, here we give an example results about Bi$_2$HMo. As shown in Fig. 4a, a distinct exchange field is induced from the Mo atoms, leading to large spin polarization of 369 meV for LCBs and 434 meV for HVBs, respectively. With SOC, local energy gaps of about 210 meV and 710 meV are opened at the K and K' points, respectively (Fig. 4b), as well as a significant valley polarization emerges in the system. The orbital-resolved band structures of $Bi_2HMo$ (Fig. S7a) indicate the bands around the $E_F$ at the two valleys are dominated by the $p_x$ and $p_y$ orbitals of Bi atoms. Fitting TB bands to the *ab-initio* results (shown in Fig. S7b), the obtained TB parameters for the ML $Bi_2HMo$ are $t_1$ = 0.60 eV, $t_2$ = 0.91 eV, 2U = 0.11 eV, $M_A$= 0.38 eV, $M_B$=0.12eV, $\lambda_{SO}$ = 0.51 eV, and $\lambda_R$ = 0.017 eV. Since the LCB at Γ point is lower than those at the two valleys, we focus on the discussion of the valley splitting of the HVBs. The valley splitting of HVBs $\Delta_{KK'}^V$ in $Bi_2HMo$ is up to 388 meV, which is very large compared to the TMD materials[21,22]. A 3880 Tesla magnetic field is needed if the valley splitting is produced by a magnetic field, since the band shift is generally of ∼0.1 meV/T by an external magnetic field[26,27] in experiments. The calculated Berry curvatures (Fig. 4b) indicate the electrons in the valleys indeed experience Lorentz-like forces. The calculation of spin Berry curvatures (Fig. S8a) and edge states (Fig. S8b) indicates $Bi_2HMo$ is in TRSB-QSH states. For the MLs $Bi_2HCr$ and $Bi_2HW$, the valley splitting of HVBs are 356 meV and 513 meV (shown in Fig. S9), respectively, both of which are record values and much larger than the previous reported maximum valley splitting[19,21,22]. With these giant valley splittings, we can readily create valley polarization with hole doping in the $Bi_2HTM$ (TM= Cr, Mo, or W) monolayers. When $E_F$ is tuned to move down in energy within the range of $\Delta_{KK'}^V$ (up to 513 meV) shown in Fig. 4b, the spin-up holes at K valley will produce a transversal current under a longitudinal in-plane electric field, giving rise to the anomalous charge/spin/valley Hall effects. Considering doping the TM atoms in the freestanding Bi monolayers may be not easy in experiments, we also explore the possibility of depositing TM atoms on the heterostructures of Bi monolayers on a SiC substrate (Bi-SiC), which has been fabricated very recently[28]. The calculated adsorption energy of Mo atom on the Bi-SiC heterostructure is about 2.0 eV and the bands of Mo@Bi-SiC are similar to those of Mo@BiH (see Fig. S10), indicating TM doping can be a realistic way to induce the spin-valley splitting in the Bi monolayers.

Besides doping TM atoms, proximity effects may be more realistic and effective way[43,44] to induce a staggered exchange field in the Bi$_2$XY system and then give rise to spin-valley polarizations. LaFeO$_3$ is a G-type antiferromagnetic (AFM) insulator with the Fe sites forming alternating (111) ferromagnetic (FM) planes[45]. Its (111) surface lattice matches well the Bi$_2$H$_2$ lattice with the mismatch of ~1%[31,46], which has been fabricated with atomic-scale control and has a very high crystallographic quality. Therefore, the LaFeO$_3$ (111) film is a very promising substrate for the ML Bi$_2$XY to produce the spin-valley splitting. A heterostructure of the ML Bi$_2$H on a (111) surface of a LaFeO$_3$ thin film is designed as displayed in Fig. S6c in Supplementary Information. The calculated large adsorption energy (3.6 eV) for the configuration indicates a very strong interaction between the ML Bi$_2$H and the substrate. Similar to the effect of doping TM atoms, the substrate also induces a staggered exchange field into the hydrogenated MLs. As shown in Fig. 4c, the Dirac bands of the ML Bi$_2$H$_2$ are strikingly spin polarized, located just inside the bulk band gap (2.1 eV) of the LaFeO$_3$ film. With the SOC considered, local band gaps of about 310 and 420 meV are opened around the K and K' points (Fig. 4d). The scale of $\Delta_{KK'}^{V}$ is found to be about 78 meV in the heterostructure, which may be tuned further by strain or an electric field. We also calculate the orbital-resolved band structures for the Bi$_2$H-LaFeO$_3$ heterostructure as shown in Fig. S7c in Supplementary Information. In this case, the HVBs at the two valleys are still dominated by the $p_x$ and $p_y$ orbitals of Bi atoms but the bands of substrate (LaFeO$_3$) exist in the nontrivial band gap from $p_x$ and $p_y$ orbitals, making the conduction bands of $p_x$ and $p_y$ orbitals higher than LCBs of the system. Fitting TB bands to the *ab-initio* results (shown in Fig. S7d), the obtained TB parameters for the Bi$_2$H-LaFeO$_3$ heterostructure are $t_1$ = 0.68 eV, $t_2$ = 1.2 eV, 2U = 0.016 eV, $M_A$ = 0.11 eV, $M_B$ = 0.03 eV, $\lambda_{SO}$ = 0.64 eV, and $\lambda_R$ = 0.012 eV. The calculation of spin Berry curvatures (Fig. S8c) and edge states (Fig. S8d) shows the Bi$_2$H-LaFeO$_3$ heterostructure is a TRSB-QSH insulator. With the hole doping, the anomalous charge/spin/valley Hall effects can also be achieved in the heterostructure.

In summary, we systematically investigated the topological properties and valleytronic behaviors in the functionalized Bi monolayers based on TB models and *ab-initio* calculations. The topologically nontrivial band gaps at the two valleys in the ML $Bi_2XY$ (X, Y = H, F, Cl, Br, or I) are found in the range from 0.891 eV to 1.256 eV. These band gaps not only give us an opportunity to generate valley polarizations with different wavelength of light, but also provide a platform for fabricating valley-light devices. Spin-valley polarizations can be generated with a staggered exchanged field introduced, which together with the staggered potential is found determining the strength of the valley splitting. The calculated spin Berry curvatures and edge states indicate TRSB-QSHEs could be observed when the $E_F$ is located in the nontrivial gap. The calculated Berry curvatures are nonzero and opposite at different valleys, driving opposite anomalous velocities of Bloch electrons. With electron (hole) doping, the spin-down electrons (spin-up holes) produce a spin-valley polarized net transverse current. Thus, with gate control, multiple Hall effects including QSHEs, anomalous charge, spin and valley Hall effects can be manipulated in the single system. Based on *ab-initio* calculations, we predict these large spin-valley polarizations and multiple Hall effects can be realized in the $Bi_2HTM$ (TM = Cr, Mo, or W) monolayers (with valley splitting of up to maximum 513 meV) or $Bi_2H/LaFeO_3$ heterostructures (with valley splitting of 78 meV). Our results not only extend the properties of known valleytronic materials, but provide new paths to realize emerging applications in electronics, spintronics, and valleytronics.

## METHODS

The geometry optimization and electronic structure calculations were performed by using the first-principles method based on density-functional theory (DFT) with the projector-augmented-wave (PAW) formalism[47], as implemented in the Vienna *ab-initio* simulation package (VASP)[48]. All calculations were carried out with a plane-wave cutoff energy of 550 eV and $12 \times 12 \times 1$ Monkhorst-Pack grids were adopted for the first Brillouin zone integral. The Berry curvatures and spin Berry Curvatures for the ML $Bi_2HF$, $Bi_2HMo$ and the $Bi_2H/LaFeO_3$ heterostructure are calculated in Wannier function bases[49]. The geometry structures and more computational details about the ML $Bi_2HMo$ and the $Bi_2H/LaFeO_3$ heterostructure are given in Supplementary Information.

**Data availability**

The data that support the findings of this study are available from the corresponding authors upon reasonable request.

## ACKNOWLEDGEMENTS

This work was supported by National Natural Science Foundation of China with Grant Nos. 11574051 and 11534001, NSF of Jiangsu Province with Grants No. BK20160007 and No. BK20170376, U.S. DOE, Office of Science BES, under Award No. DE-SC0004890 (I. Z.), Fudan High-end Computing Center and UB Center for Computational Research.

## AUTHOR CONTRIBUTIONS

Z.Y. conceived the research. T.Z. carried out the calculations. T.Z., I.Z., and Z.Y. wrote the manuscript. J.Z. and H.J. contributed to the discussion and editing of the manuscript. I.Z. and Z.Y. are responsible for coordinating the project.

## ADDITIONAL INFORMATION

**Supplementary Information Available:** The band gaps and band structures for the ML $Bi_2XY$ (X, Y = H, F, Cl, Br, or I) monolayers, orbital-resolved DFT bands and fitted bands from the TB model for the ML $Bi_2HF$, ML $Bi_2HMo$, and $Bi_2H/LaFeO_3$ heterostructure, bands of Bi-SiC heterostructure with Mo deposited, the geometry structures, electronic structures, energy levels at the two valleys, and more computational details of the ML $Bi_2HTM$ (TM = Cr, Mo, or W) and the $Bi_2H/LaFeO_3$ heterostructure are shown in Supplementary Information. Some details of the calculations of TB model are also given in Supplementary Information.

**Competing interests:** The authors declare no competing interests.

## REFERENCES

1. Xiao, D., Yao, W. & Niu, Q. Valley-Contrasting Physics in Graphene: Magnetic Moment and Topological Transport. *Phys. Rev. Lett.* **99**, 236809 (2007).
2. Schaibley, J. R._, et al._ Valleytronics in 2D materials. *Nat. Rev. Mater.* **1**, 16055 (2016).
3. Xu, X., Yao, W., Xiao, D. & Heinz, T. F. Spin and pseudospins in layered transition metal dichalcogenides. *Nat. Phys.* **10**, 343-350 (2014).

4. Mak, K. F., McGill, K. L., Park, J. & McEuen, P. L. The valley Hall effect in MoS$_2$ transistors. *Science* **344**, 1489-1492 (2014).

5. Takashina, K.*, et al.* Valley Polarization in Si(100) at Zero Magnetic Field. *Phys. Rev. Lett.* **96**, 236801 (2006).

6. Isberg, J.*, et al.* Generation, transport and detection of valley-polarized electrons in diamond. *Nat. Mater.* **12**, 760-764 (2013).

7. Shkolnikov, Y. P., De Poortere, E. P., Tutuc, E. & Shayegan, M. Valley Splitting of AlAs Two-Dimensional Electrons in a Perpendicular Magnetic Field. *Phys. Rev. Lett.* **89**, 226805 (2002).

8. Yao, W., Xiao, D. & Niu, Q. Valley-dependent optoelectronics from inversion symmetry breaking. *Phys. Rev. B* **77**, 235406 (2008).

9. Zhang, F.*, et al.* Spontaneous Quantum Hall States in Chirally Stacked Few-Layer Graphene Systems. *Phys. Rev. Lett.* **106**, 156801 (2011).

10. Cheng, S., Zhou, J., Jiang, H. & Sun, Q.-F. The valley filter efficiency of monolayer graphene and bilayer graphene line defect model. *New J. Phys.* **18**, 103024 (2016).

11. Gorbachev, R. V.*, et al.* Detecting topological currents in graphene superlattices. *Science* **346**, 448-451 (2014).

12. Žutić, I., Fabian, J. & Das Sarma, S. Spintronics: Fundamentals and applications. *Rev. Mod. Phys.* **76**, 323-410 (2004).

13. Ang, Y. S.*, et al.* Valleytronics in merging Dirac cones: All-electric-controlled valley filter, valve, and universal reversible logic gate. *Phys. Rev. B* **96**, 245410 (2017).

14. Feng, W.*, et al.* Intrinsic spin Hall effect in monolayers of group-VI dichalcogenides: A first-principles study. *Phys. Rev. B* **86**, 165108 (2012).

15. Tong, W.-Y., Gong, S.-J., Wan, X. & Duan, C.-G. Concepts of ferrovalley material and anomalous valley Hall effect. *Nat. Commun.* **7**, 13612 (2016).

16. Nirpendra, S. & Udo, S. A Route to Permanent Valley Polarization in Monolayer MoS$_2$. *Adv. Mater.* **29**, 1600970 (2017).

17. Cheng, Y. C., Zhang, Q. Y. & Schwingenschlögl, U. Valley polarization in magnetically doped single-layer transition-metal dichalcogenides. *Phys. Rev. B* **89**, 155429 (2014).

18. Chen, X., Zhong, L., Li, X. & Qi, J. Valley splitting in the transition-metal dichalcogenide monolayer via atom adsorption. *Nanoscale* **9**, 2188-2194 (2017).

19. Qi, J., Li, X., Niu, Q. & Feng, J. Giant and tunable valley degeneracy splitting in MoTe$_2$. *Phys. Rev. B* **92**, 121403 (2015).

20. Zhang, Q., *et al.* Large Spin-Valley Polarization in Monolayer MoTe$_2$ on Top of EuO(111). *Adv. Mater.* **28**, 959-966 (2016).

21. Li, N., *et al.* Large valley polarization in monolayer MoTe$_2$ on a magnetic substrate. *Phys. Chem. Chem. Phys.* **20**, 3805-3812 (2018).

22. Zhao, C., *et al.* Enhanced valley splitting in monolayer WSe$_2$ due to magnetic exchange field. *Nat. Nanotechnol.* **12**, 757-762 (2017).

23. Cao, T., *et al.* Valley-selective circular dichroism of monolayer molybdenum disulphide. *Nat. Commun.* **3**, 887 (2012).

24. Zeng, H., *et al.* Valley polarization in MoS$_2$ monolayers by optical pumping. *Nat. Nanotechnol.* **7**, 490-493 (2012).

25. Mak, K. F., He, K., Shan, J. & Heinz, T. F. Control of valley polarization in monolayer MoS$_2$ by optical helicity. *Nat. Nanotechnol.* **7**, 494-498 (2012).

26. MacNeill, D., *et al.* Breaking of Valley Degeneracy by Magnetic Field in Monolayer MoSe$_2$. *Phys. Rev. Lett.* **114**, 037401 (2015).

27. Aivazian, G., *et al.* Magnetic control of valley pseudospin in monolayer WSe$_2$. *Nat. Phys.* **11**, 148-152 (2015).

28. Reis, F., *et al.* Bismuthene on a SiC substrate: A candidate for a high-temperature quantum spin Hall material. *Science* **357**, 287-290 (2017).

29. Dominguez, F., *et al.* Testing Topological Protection of Edge States in Hexagonal Quantum Spin Hall Candidate Materials. Preprint at arXiv:1803.02648 (2018).

30. Jin, K.-H. & Jhi, S.-H. Quantum anomalous Hall and quantum spin-Hall phases in flattened Bi and Sb bilayers. *Sci. Rep.* **5**, 8426 (2015).

31. Song, Z., *et al.* Quantum spin Hall insulators and quantum valley Hall insulators of BiX/SbX (X=H, F, Cl and Br) monolayers with a record bulk band gap. *NPG Asia Mater.* **6**, e147 (2014).

32. Ma, Y., *et al.* Robust Two-Dimensional Topological Insulators in Methyl-Functionalized Bismuth, Antimony, and Lead Bilayer Films. *Nano Lett.* **15**, 1083-1089 (2015).

33. Zhou, T., *et al.* Quantum spin-quantum anomalous Hall effect with tunable edge states in Sb monolayer-based heterostructures. *Phys. Rev. B* **94**, 235449 (2016).

34. Qiao, Z., *et al.* Quantum anomalous Hall effect in graphene from Rashba and exchange effects. *Phys. Rev. B* **82**, 161414 (2010).

35. Qiao, Z., *et al.* Microscopic theory of quantum anomalous Hall effect in graphene. *Phys. Rev. B* **85**, 115439 (2012).

36. Yang, W., *et al.* Electrically Tunable Valley-Light Emitting Diode (vLED) Based on CVD-Grown Monolayer WS$_2$. *Nano Lett.* **16**, 1560-1567 (2016).

37. Yao, Y., *et al.* First Principles Calculation of Anomalous Hall Conductivity in Ferromagnetic bcc Fe. *Phys. Rev. Lett.* **92**, 037204 (2004).

38. Scharf, B., Xu, G., Matos-Abiague, A. & Žutić, I. Magnetic Proximity Effects in Transition-Metal Dichalcogenides: Converting Excitons. *Phys. Rev. Lett.* **119**, 127403 (2017).

39. Song, Z., *et al.* Spontaneous Valley Splitting and Valley Pseudospin Field Effect Transistor of Monolayer $VAgP_2Se_6$. Preprint at arXiv:1801.03173 (2018).

40. Ezawa, M. Spin valleytronics in silicene: Quantum spin Hall–quantum anomalous Hall insulators and single-valley semimetals. *Phys. Rev. B* **87**, 155415 (2013).

41. Zhou, T., *et al.* Quantum Spin-Quantum Anomalous Hall Insulators and Topological Transitions in Functionalized Sb(111) Monolayers. *Nano Lett.* **15**, 5149-5155 (2015).

42. Yang, Y., *et al.* Time-Reversal-Symmetry-Broken Quantum Spin Hall Effect. *Phys. Rev. Lett.* **107**, 066602 (2011).

43. Qiao, Z., *et al.* Quantum Anomalous Hall Effect in Graphene Proximity Coupled to an Antiferromagnetic Insulator. *Phys. Rev. Lett.* **112**, 116404 (2014).

44. Žutić, I. et al. Proximitized materials. Materials Today (in press), https://doi.org/10.1016/j.mattod.2018.05.003 (2018).

45. Seo, J. W. et al. Antiferromagnetic $LaFeO_3$ thin films and their effect on exchange bias. J. Phys. Condens. Matter **20**, 264014 (2008).

46. Ueda, K., Tabata, H. & Kawai, T. Control of magnetic properties in $LaCrO_3$–$LaFeO_3$ artificial superlattices. *Science* **280**, 1064-1066 (1998).

47. Kresse, G. & Joubert, D. From ultrasoft pseudopotentials to the projector augmented-wave method. *Phys. Rev. B* **59**, 1758-1775 (1999).

48. Kresse, G. & Furthmüller, J. Efficient iterative schemes for *ab initio* total-energy calculations using a plane-wave basis set. *Phys. Rev. B* **54**, 11169-11186 (1996).

49. Mostofi, A. A., *et al.* wannier90: A tool for obtaining maximally-localised Wannier functions. *Comput. Phys. Commun.* **178**, 685-699 (2008).

**Figures and captions**

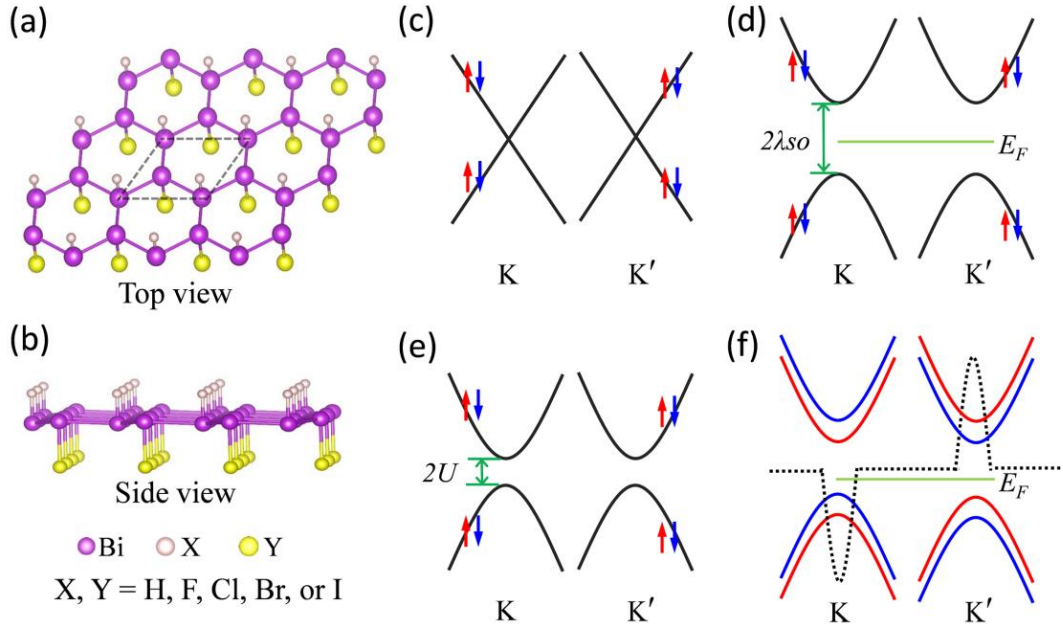

**Fig. 1**. Geometry and band structures of functionalized Bi monolayers without magnetization. (a) Top and (b) side views of the structure of ML $Bi_2XY$ (X, Y = H, F, Cl, Br, or I). (c) and (d) Energy bands for the ML $Bi_2XY$ (X = Y) system from the TB model without and with SOC considered, respectively. (e) and (f) The same as (c) and (d), respectively, except for X ≠ Y. The small red/blue arrow in (c)-(d) indicates the spin-up/spin-down state. The red/blue curve in (f) indicates the spin-up/spin-down state. The black dots in (f) denote the Berry curvatures for the whole valence bands. The TB parameters of $t_1$ = 0.8 eV, $t_2$ = 1.0 eV, $2U$ = 0.0/0.12 eV, $\lambda_{SO}$ = 0.6 eV, and $\lambda_R$ = 0.015 eV are adopted to plot (c)-(d)/(e)-(f).

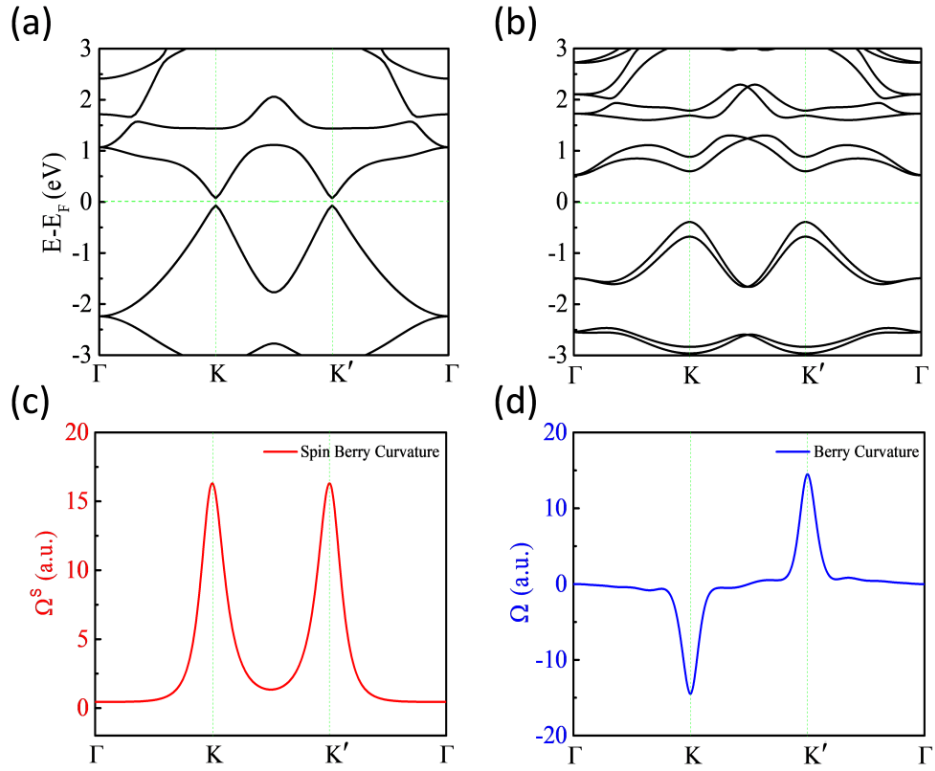

**Fig. 2**. Band structures and topological properties of the ML $Bi_2HF$. (a) and (b) Band structures for the ML $Bi_2HF$ without and with SOC, respectively. (c) and (d) Spin Berry curvatures and Berry curvatures for the whole valence bands of (b). The results are obtained based on density functional theory calculations.

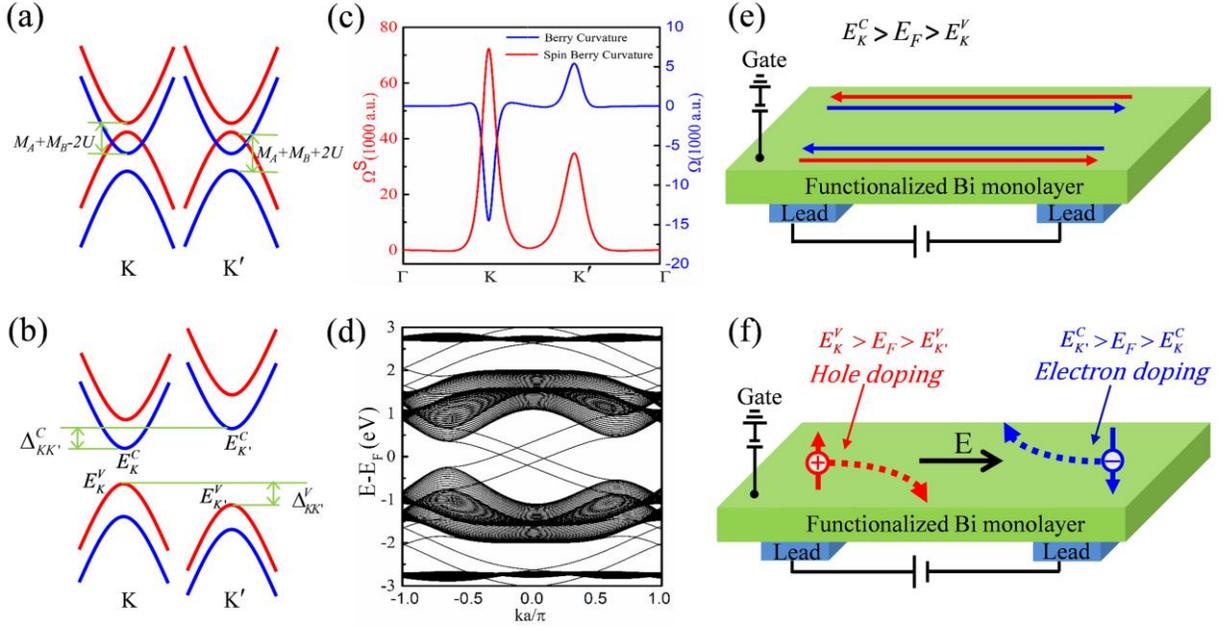

**Fig. 3**. Spin-valley polarization and multiple Hall effect in functionalized Bi monolayers. (a) and (b) Energy bands for the ML $Bi_2XY$ from the TB model with a staggered exchange field ($M_B < M_A < \lambda_{SO}$) without and with SOC, respectively. The red/blue curves indicate the spin-up/spin-down states. (c) Spin Berry curvatures and Berry curvatures for the whole valence bands of (b). (d) The band structures of the zigzag nanoribbon (containing 20 zigzag chains in the width) for the system of (b). (e) Schematic of the QSHE. The red/blue arrows indicate the spin-up/down edge states. (f) Schematic of anomalous charge/spin/valley Hall effects with hole or electron doping. The electrons and holes are indicated by circles with −/+, respectively. The TB parameters of $t_1 = 0.8$ eV, $t_2 = 1.0$ eV, $2U = 0.12$ eV, $M_A = 0.4$ eV, and $M_B = 0.1$ eV, $\lambda_{SO} = 0.0/0.6$ eV, and $\lambda_R = 0.0/0.015$ eV are adopted to plot (a)/(b)-(d).

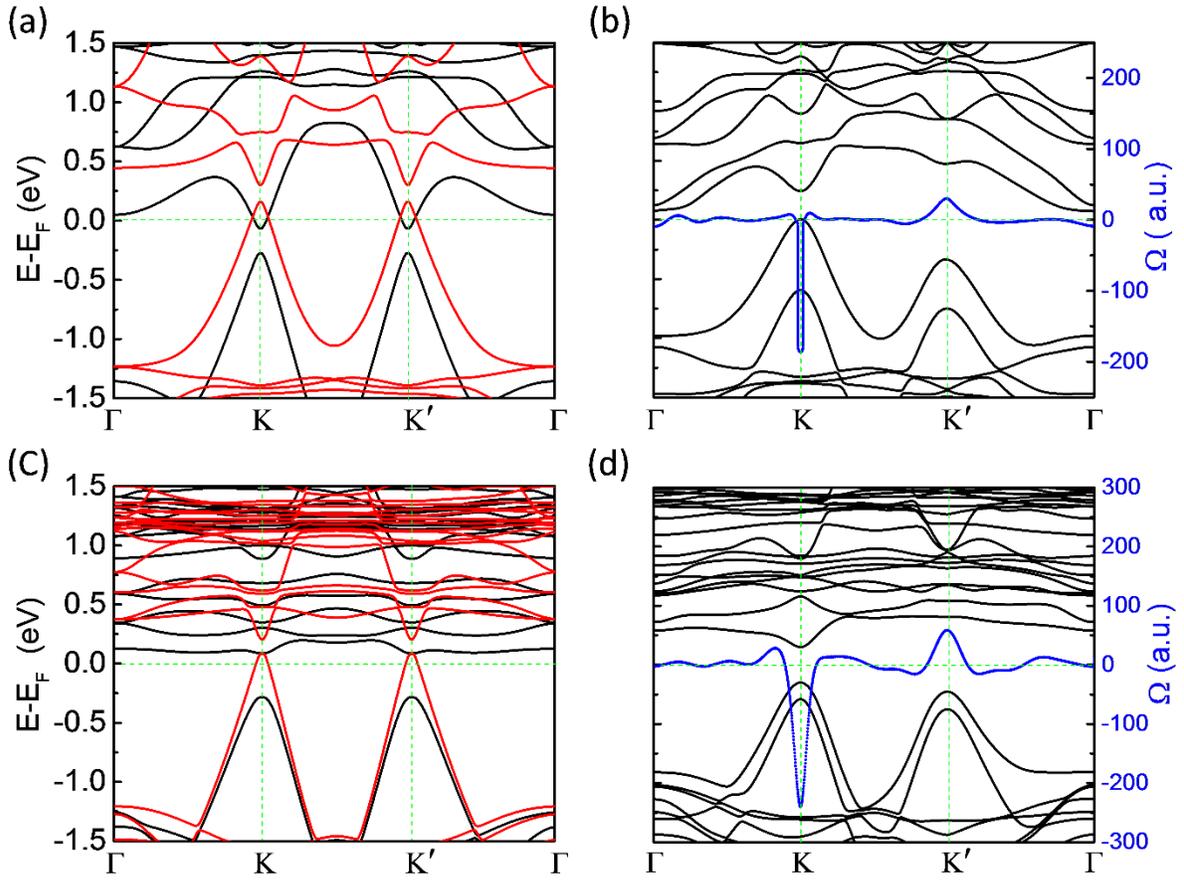

**Fig. 4**. Band structures and topological properties of magnetic functionalized Bi monolayers. (a) and (b) Band structures for the functionalized Bi MLs with H atoms deposited on the upper surface and Mo atoms deposited on the lower surface (Bi$_2$HMo) without and with SOC, respectively. (c) and (d) Band structures for the Bi$_2$H/LaFeO$_3$ heterostructure without and with SOC, respectively. The red and black curves in (a) and (c) denote the spin-up and spin-down states, respectively. The blue color in (b) and (d) denotes the calculated Berry curvatures for the whole valence bands. The results are obtained based on density functional theory calculations.

Supplementary Information

# Giant Spin-valley Polarization and Multiple Hall Effect in Functionalized Bi Monolayers

*Tong Zhou[1,2], Jiayong Zhang[1,3], Hua Jiang[4], Igor Zutic[2], and Zhongqin Yang[1,5,\*]*

[1]State Key Laboratory of Surface Physics and Key Laboratory for Computational Physical Sciences (MOE) & Department of Physics, Fudan University, Shanghai 200433, China

[2]Department of Physics, University at Buffalo, SUNY, Buffalo, NY 14260, USA

[3]Jiangsu Key Laboratory of Micro and Nano Heat Fluid Flow Technology and Energy Application, School of Mathematics and Physics, Suzhou University of Science and Technology, Suzhou, Jiangsu, China

[4]College of Physics, Optoelectronics and Energy Soochow University, Suzhou (215006), China

[5]Collaborative Innovation Center of Advanced Microstructures, Fudan University, Shanghai 200433, China

*Address correspondence to: zyang@fudan.edu.cn

**Table S1** Band gaps (eV) at the two valleys for the Bi$_2$XY (X, Y = H, F, Cl, Br, or I) monolayers with spin-orbit coupling (SOC).

| X \ Y | H | F | Cl | Br | I |
|---|---|---|---|---|---|
| H | 1.254 | 1.011 | 1.224 | 1.231 | 1.210 |
| F | 1.011 | 1.160 | 1.019 | 0.969 | 0.891 |
| Cl | 1.224 | 1.019 | 1.306 | 1.256 | 1.171 |
| Br | 1.231 | 0.969 | 1.256 | 1.301 | 1.218 |
| I | 1.210 | 0.891 | 1.171 | 1.218 | 1.297 |

**Table S2** Band gaps (eV) at the two valleys for the Bi$_2$XY (X, Y = H, F, Cl, Br, or I) monolayers without SOC.

| X \ Y | H | F | Cl | Br | I |
|---|---|---|---|---|---|
| H | 0 | 0.141 | 0.006 | 0.025 | 0.039 |
| F | 0.141 | 0 | 0.196 | 0.236 | 0.300 |
| Cl | 0.006 | 0.196 | 0 | 0.038 | 0.096 |
| Br | 0.025 | 0.236 | 0.038 | 0 | 0.058 |
| I | 0.039 | 0.300 | 0.096 | 0.058 | 0 |

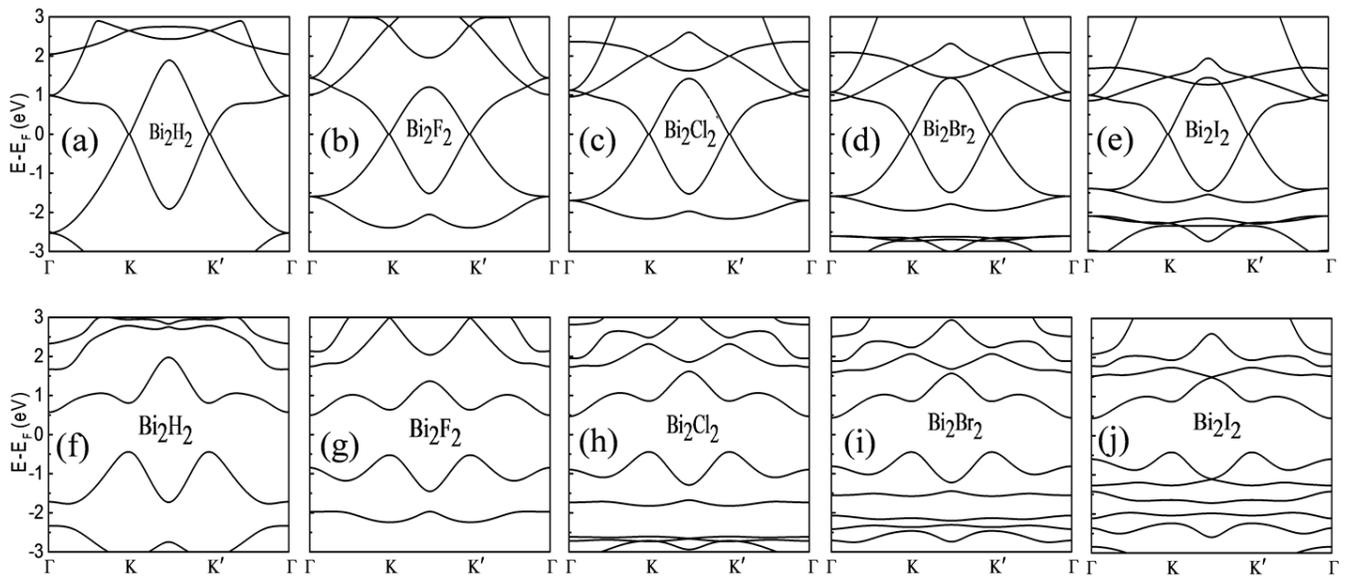

**Fig. S1**. (a) – (e) Band structures for the $Bi_2X_2$ (X = H, F, Cl, Br, or I) monolayers without SOC. (f) – (j) Band structures for the $Bi_2X_2$ (X = H, F, Cl, Br, or I) monolayers with SOC.

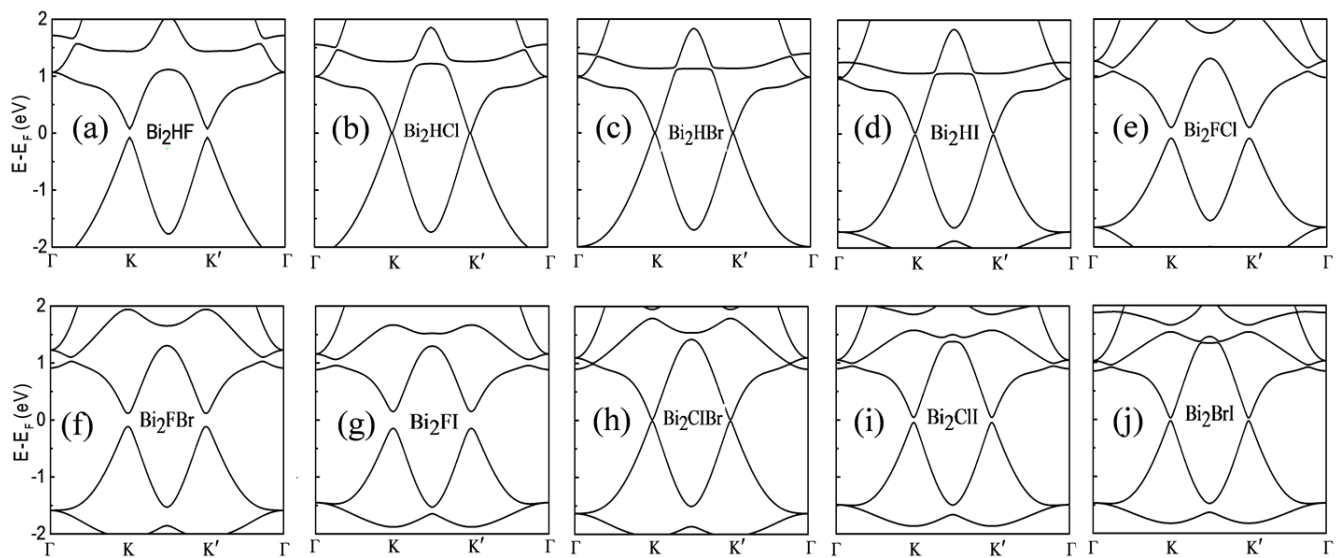

**Fig. S2**. (a) – (j) Band structures for the Bi$_2$XY (X, Y = H, F, Cl, Br, or I and X≠Y) monolayers without SOC.

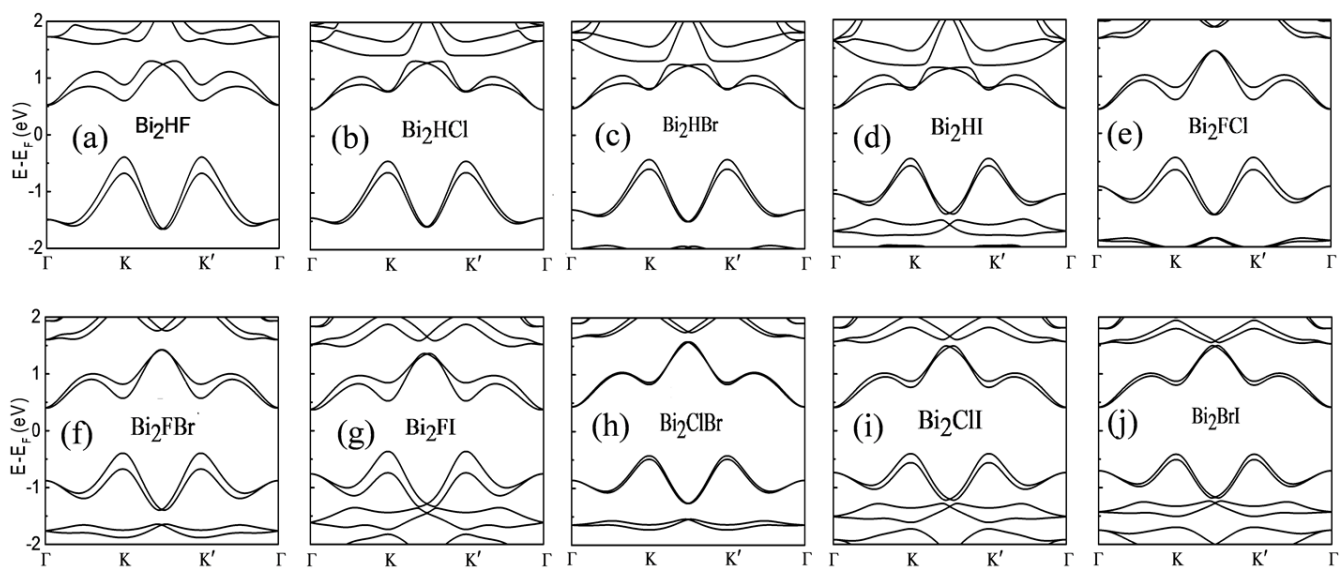

**Fig. S3**. (a) – (j) Band structures for the Bi$_2$XY (X, Y = H, F, Cl, Br, or I and X≠Y) monolayers with SOC.

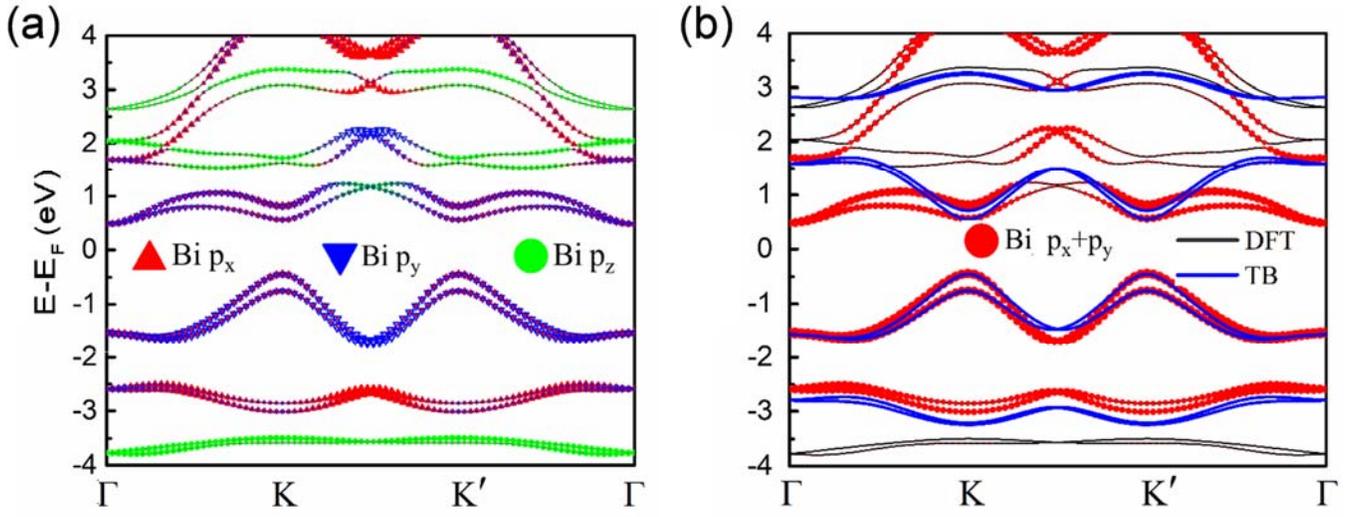

**Fig. S4**. (a) Orbital-resolved band structures for the Bi$_2$HF monolayer with SOC, obtained from the DFT calculations. (b) Fitted bands of the TB model. The results from the DFT calculations are also shown. The TB parameters for the Bi$_2$HF monolayer are $t_1$ = 0.73 eV, $t_2$ = 1.06 eV, $2U$ = 0.14 eV, $\lambda_{SO}$ = 0.61 eV, and $\lambda_R$ = 0.026 eV.

## Spin-valley splittings from the TB model

Here we analyze the energy levels of the bands at the two valleys. At K point with $k_x = 0$, $k_y = \frac{4\pi}{3\sqrt{3}}$ ( the lattice parameter of the unit cell in our TB model is assumed as 1), $H_3$ in k-space at the K point is written as

$$H_{3K} = \begin{bmatrix} U+M_A+\lambda so & 0 & 0 & 0 & 0 & 0 & 3t_2 & 0 \\ 0 & U-M_A-\lambda so & 0 & 0 & 0 & 0 & 0 & 3t_2 \\ 0 & 0 & U+M_A-\lambda so & 0 & 0 & 0 & 0 & 0 \\ 0 & 0 & 0 & U-M_A+\lambda so & 0 & 0 & 0 & 0 \\ 0 & 0 & 0 & 0 & -U+M_B+\lambda so & 0 & 0 & 0 \\ 0 & 0 & 0 & 0 & 0 & -U-M_B-\lambda so & 0 & 0 \\ 3t_2 & 0 & 0 & 0 & 0 & 0 & -U+M_B-\lambda so & 0 \\ 0 & 3t_2 & 0 & 0 & 0 & 0 & 0 & -U-M_B+\lambda so \end{bmatrix}.$$

The energy levels at this K point can be obtained by diagonalizing above matrix. Around the $E_F$, the energy levels can be analytically expressed as $E_1 = \lambda so + M_B - U$, $E_2 = \lambda so - M_A + U$, $E_3 = -\lambda so + M_A + U$, and $E_4 = -\lambda so - M_B - U$. Similarly, at the K' point with $k_x = 0$, $k_y = \frac{8\pi}{3\sqrt{3}}$,

$$H_{3K'} = \begin{bmatrix} U+M_A+\lambda so & 0 & 0 & 0 & 0 & 0 & 0 & 0 \\ 0 & U-M_A-\lambda so & 0 & 0 & 0 & 0 & 0 & 0 \\ 0 & 0 & U+M_A-\lambda so & 0 & 3t_2 & 0 & 0 & 0 \\ 0 & 0 & 0 & U-M_A+\lambda so & 0 & 3t_2 & 0 & 0 \\ 0 & 0 & 3t_2 & 0 & -U+M_B+\lambda so & 0 & 0 & 0 \\ 0 & 0 & 0 & 3t_2 & 0 & -U-M_B-\lambda so & 0 & 0 \\ 0 & 0 & 0 & 0 & 0 & 0 & -U+M_B-\lambda so & 0 \\ 0 & 0 & 0 & 0 & 0 & 0 & 0 & -U-M_B+\lambda so \end{bmatrix}.$$

The energy levels near the $E_F$ are $E_1' = \lambda so + M_A + U$, $E_2' = \lambda so - M_B - U$, $E_3' = -\lambda so + M_B - U$, $E_4' = -\lambda so - M_A + U$. Since the SOC of Bi atoms, primarily coming from the $p_x$ and $p_y$ orbitals, is very large, $\lambda_{SO} > M_A > M_B$ can be assumed in the TB model calculations. In addition, the induced magnetic field is probably larger than the staggered potential ($M_A + M_B > 2U$). Under these conditions, the sequences of the energy levels of $H_3$ around the Fermi level, $E_F$, at the two valleys are obtained as Fig. S4. Here we also give the Rashba term as $H_R = \sum_{i \in A} \sum_{\delta=1,2,3} \Phi_i^+ T_{R\delta} \Phi_{i+\delta}$ for the functionalized Bi monolayers, $H_R$ in k-space can be written as

$$H_R = T_{R\delta_1} e^{ik_x a} + T_{R\delta_2} e^{-i(k_x a/2 - k_y \sqrt{3}a/2)} + T_{R\delta_3} e^{-i(k_x a/2 + k_y \sqrt{3}a/2)}, \text{ where}$$

$$T_{R\delta_1} = -i \begin{bmatrix} \lambda_R & \lambda_R' \\ \lambda_R' & \lambda_R \end{bmatrix} \otimes \sigma_y, \quad T_{R\delta_2} = i \begin{bmatrix} \lambda_R & z^2\lambda_R' \\ z\lambda_R' & \lambda_R \end{bmatrix} \otimes (\frac{\sqrt{3}}{2}\sigma_x + \frac{1}{2}\sigma_y), \quad T_{R\delta_3} = i \begin{bmatrix} \lambda_R & z\lambda_R' \\ z^2\lambda_R' & \lambda_R \end{bmatrix} \otimes (-\frac{\sqrt{3}}{2}\sigma_x + \frac{1}{2}\sigma_y).$$

Because the Rashba term is much smaller than the $M_A(M_B)$, $\lambda_{SO}$, and $U$ terms in the functionalized Bi monolayers, we neglect the Rashba SOC in $H_3$.

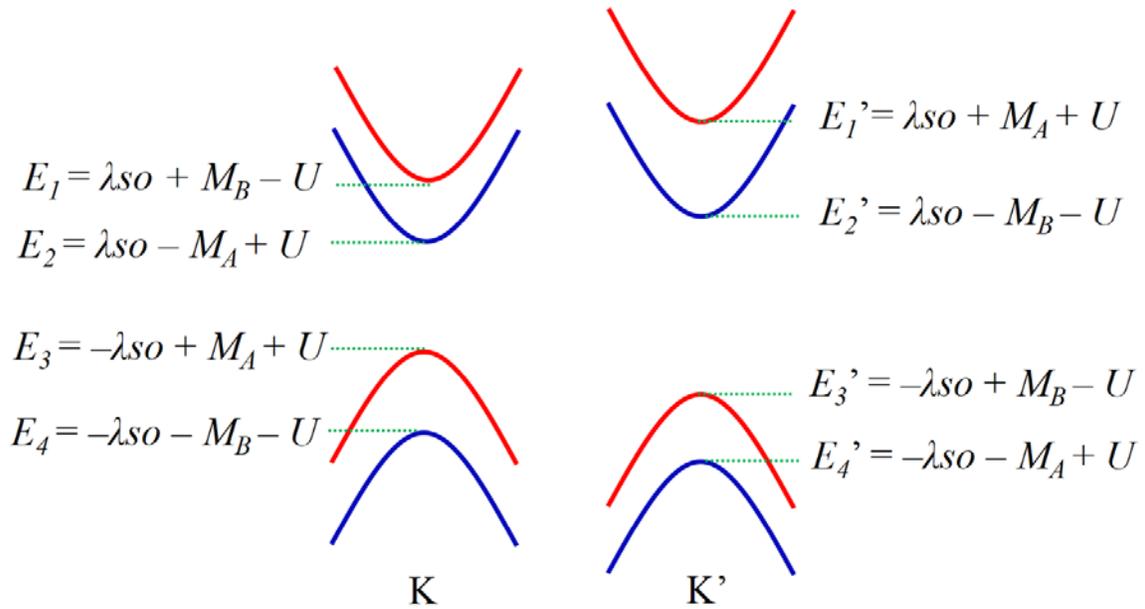

**Fig. S5** Schematic depictions of the energy levels of the bands at the two valleys for $H_3$ when $M_B < M_A < \lambda_{SO}$ and $M_B + M_A > 2U$. The red/blue curves indicate the spin-up/spin-down states.

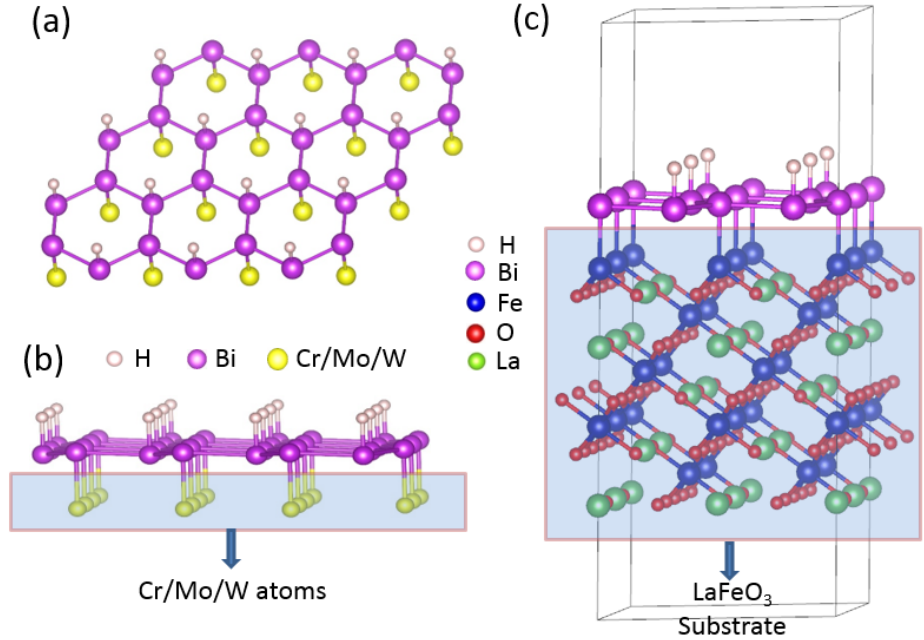

**Fig. S6**. (Color online) (a) Top view of the Bi$_2$HTm (Tm= Cr, Mo, or W) monolayer. (b) Side view of the Bi$_2$HTm monolayer. (c) Structure of the Bi$_2$H/LaFeO$_3$ heterostructure. The geometry optimization and electronic structure calculations were performed by using the first-principles method based on density-functional theory (DFT) with the projector-augmented-wave (PAW) formalism[1], as implemented in the Vienna ab-initio simulation package (VASP)[2]. All calculations were carried out with a plane-wave cutoff energy of 550 eV and 12 × 12 × 1 Monkhorst-Pack grids were adopted for the first Brillouin zone integral. For the Bi$_2$HTm monolayer, the Perdew-Burke-Ernzerhof generalized-gradient approximation (GGA) was used to describe the exchange and correlation functional[3]. The optimized lattice constants are found to be 5.53 Å for all of the three structures. For the Bi$_2$H/LaFeO$_3$ heterostructure, the substrates with different thicknesses containing four to seven Fe layers are chosen (there are five Fe layers in (c)). It was found that they gave the same results. To stabilize the substrate, the bottom Fe terminal surface of the LaFeO$_3$ film, far away from the Bi monolayer interface, was passivated by two fluorine atoms per unit cell. A vacuum space of larger than 15 Å was used to avoid the interaction between two adjacent heterostructure slabs. For the structural relaxation, the Bi$_2$H sheet and the topmost Fe, La, and O atoms in the substrate were allowed to relax until the Hellmann-Feynman force on each atom was smaller than 0.01 eV/Å. To take into account the correlation effects of Fe 3$d$ electrons, the GGA+$U$ method was adopted and the value of the effective Hubbard $U_{eff} = U - J$ was chosen to be 3.5 eV, which can correctly describe the correlation effect of Fe 3$d$ electrons in LaFeO$_3$ bulk materials[4]. The van der Waals (vdW) interaction correction with Grimme (DFT-D2)[5] method was considered in the calculations, including the structural relaxation.

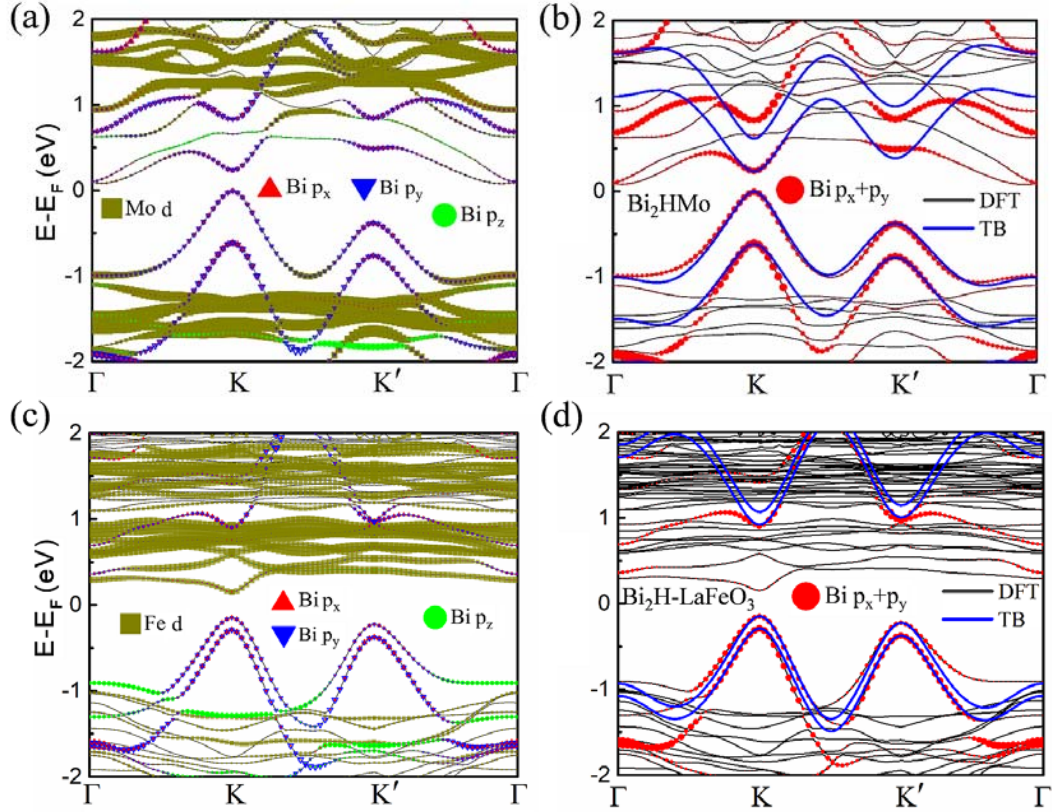

**Fig. S7**. (a) and (c) Orbital-resolved band structures with SOC for the $Bi_2HMo$ monolayer and the $Bi_2H$-$LaFeO_3$ heterostructure, respectively, obtained from the DFT calculations. (b) and (d) The fitted bands of the TB model for the $Bi_2HMo$ monolayer and the $Bi_2H$-$LaFeO_3$ heterostructure, respectively. The DFT results are also shown. The TB parameters obtained for the $Bi_2HMo$ monolayer are $t_1 = 0.60$ eV, $t_2 = 0.91$ eV, $2U = 0.11$ eV, $M_A= 0.38$ eV, $M_B=0.12$ eV, $\lambda_{SO} = 0.51$ eV, and $\lambda_R = 0.017$ eV. The TB parameters obtained for the $Bi_2H$-$LaFeO_3$ heterostructure are $t_1 = 0.68$ eV, $t_2 = 1.2$ eV, $2U = 0.016$ eV, $M_A= 0.11$ eV, $M_B=0.03$ eV, $\lambda_{SO} = 0.64$ eV, and $\lambda_R = 0.012$ eV.

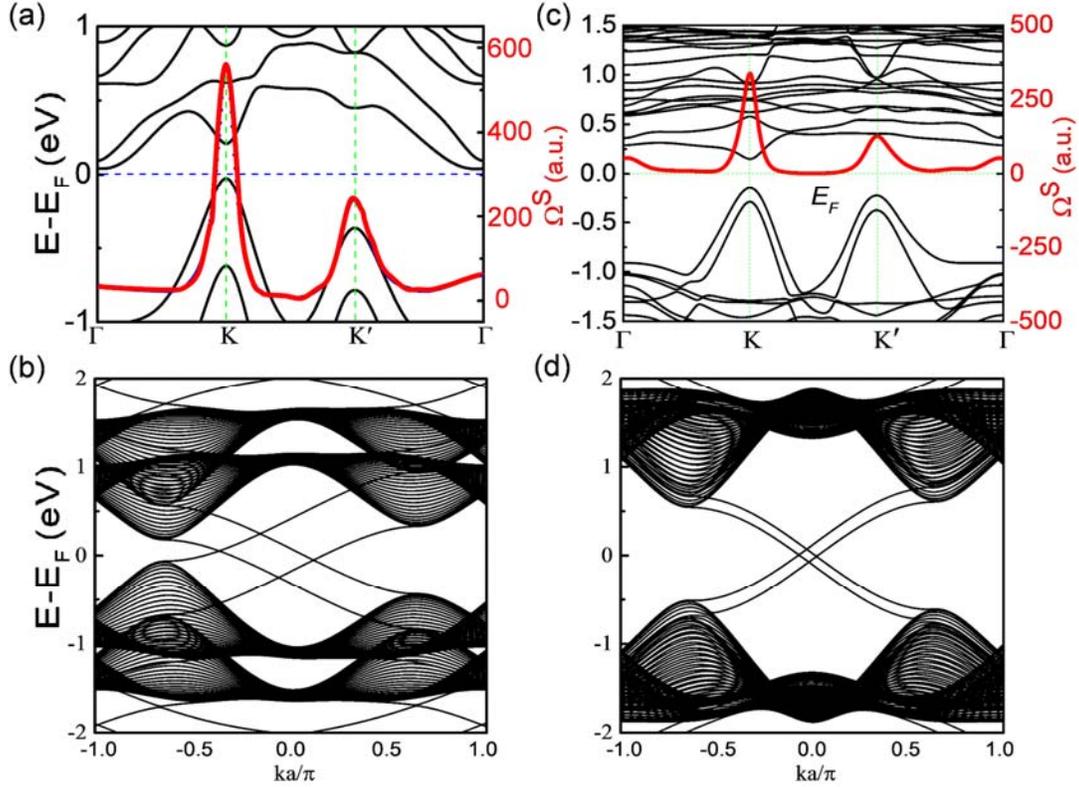

**Fig. S8** (a) Bands and spin Berry curvature for the whole valence bands for the ML $Bi_2HMo$. (b) The band structure of the zigzag nanoribbon (containing 20 zigzag chains in the width) from the ML $Bi_2HMo$, obtained with TB parameters of $t_1 = 0.68$ eV, $t_2 = 1.2$ eV, $2U = 0.016$ eV, $M_A = 0.11$ eV, $M_B = 0.03$ eV, $\lambda_{SO} = 0.64$ eV, and $\lambda_R = 0.012$ eV. (c) Bands and spin Berry curvature for the whole valence bands for the $Bi_2H$-$LaFeO_3$ heterostructure (d) The band structures of the zigzag nanoribbon (containing 20 zigzag chains in the width) from the $Bi_2H$-$LaFeO_3$ heterostructure, obtained with TB parameters of $t_1 = 0.68$ eV, $t_2 = 1.2$ eV, $2U = 0.016$ eV, $M_A = 0.11$ eV, $M_B = 0.03$ eV, $\lambda_{SO} = 0.64$ eV, and $\lambda_R = 0.012$ eV.

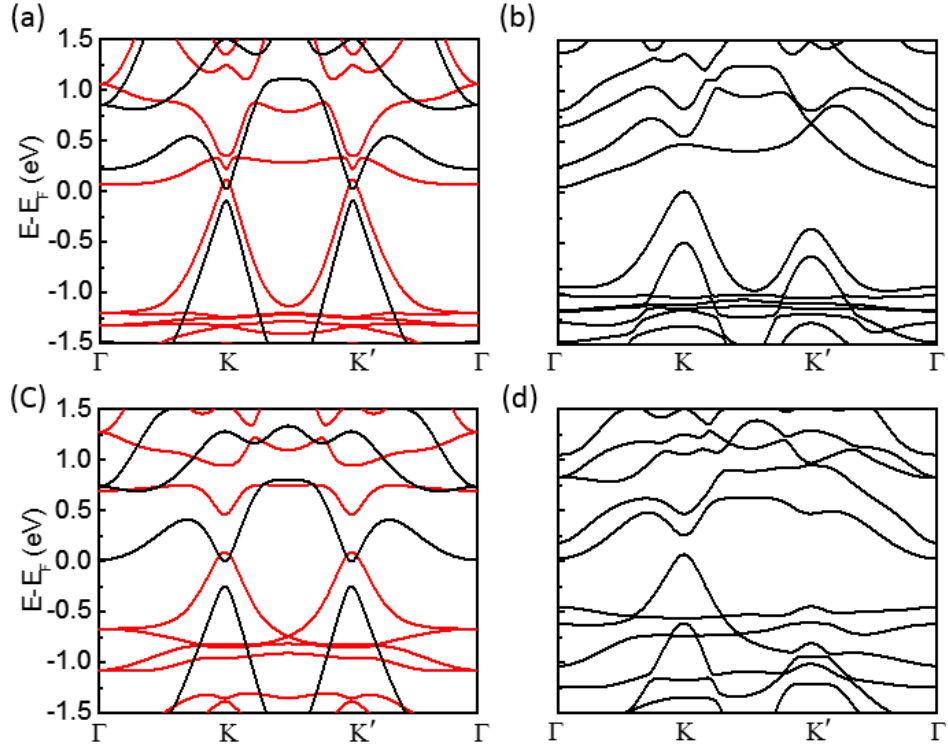

**Fig. S9** (a) and (b) Band structures for the Bi$_2$HCr monolayer without and with SOC considered, respectively. (c) and (d) Band structures for the Bi$_2$HW monolayer without and with SOC considered, respectively. The valley splittings of the highest valence bands in the Bi$_2$HCr and Bi$_2$HW monolayers are 356 meV and 513 meV, respectively. The red and black curves in (a) and (c) indicate spin-up and down states, respectively.

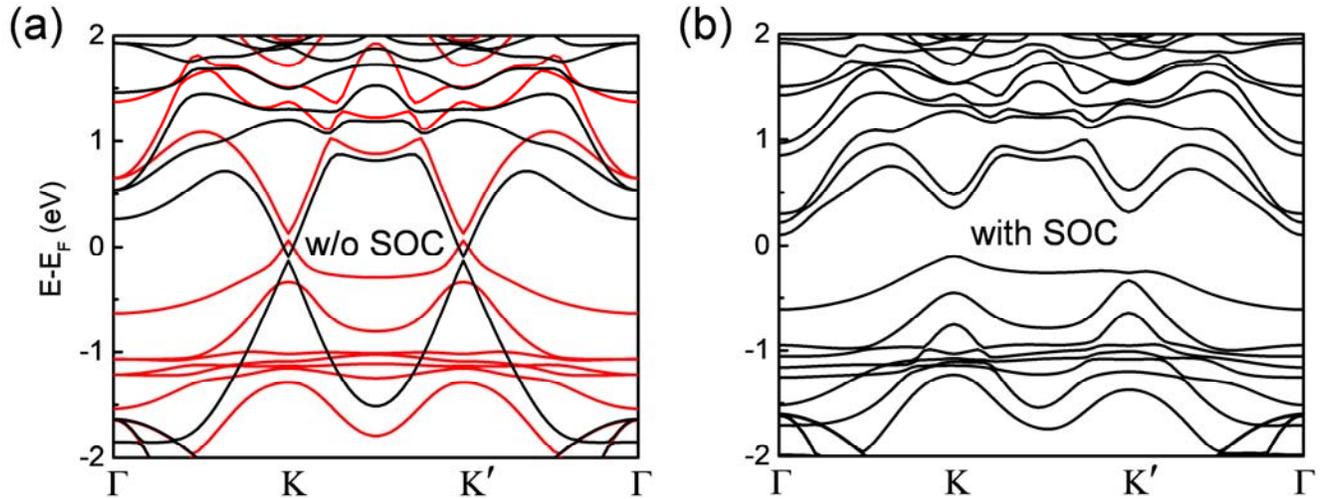

**Fig. S10** (a) and (b) Band structures for the Mo@Bi-SiC heterostructure without and with SOC considered, respectively. The red and black curves in (a) indicate spin-up and down states, respectively.

**REFERENCES**

1. Kresse, G. & Joubert, D. From ultrasoft pseudopotentials to the projector augmented-wave method. *Phys. Rev. B* **59**, 1758-1775 (1999).
2. Kresse, G. & Furthmüller, J. Efficient iterative schemes for ab initio total-energy calculations using a plane-wave basis set. *Phys. Rev. B* **54**, 11169-11186 (1996).
3. Perdew, J. P., Burke, K. & Ernzerhof, M. Generalized gradient approximation made simple. *Phys. Rev. Lett.* **77**, 3865-3868 (1996).
4. Arima, T., Tokura, Y. & Torrance, J. Variation of optical gaps in perovskite-type 3d transition-metal oxides. *Phys. Rev. B* **48**, 17006-17009 (1993).
5. Grimme, S. Semiempirical GGA−type density functional constructed with a long−range dispersion correction. *J. Comput. Chem.* **27**, 1787-1799 (2006).